# Circularly polarized microwaves for magnetic resonance study in the GHz range: application to nitrogen-vacancy in diamonds

M. Mrózek,[1,a)] J. Mlynarczyk,[2] D.S. Rudnicki,[1] and W. Gawlik[1]

[1]*Institute of Physics, Jagiellonian University, Lojasiewicza 11, 30-348 Krakow, Poland*
[2]*Department of Electronics, AGH University of Science and Technology, Krakow, Poland*

The ability to create time-dependent magnetic fields of controlled polarization is essential for many experiments with magnetic resonance. We describe a microstrip circuit that allows us to generate strong magnetic field at microwave frequencies with arbitrary adjusted polarization. The circuit performance is demonstrated by applying it to an optically detected magnetic resonance and Rabi nutation experiments in nitrogen-vacancy color centers in diamond. Thanks to high efficiency of the proposed microstrip circuit and degree of circular polarization of 85% it is possible to address the specific spin states of a diamond sample using a low power microwave generator. The circuit may be applied to a wide range of magnetic resonance experiments with a well-controlled polarization of microwaves.

Creation of time-dependent magnetic fields is essential for experiments and applications of magnetic resonance phenomena, such as nuclear magnetic resonance (NMR), magnetic resonance imaging (MRI), electron spin resonance (ESR), etc. In most such applications **linearly polarized** fields are used. In many cases, however, it is desirable to create **circularly polarized** fields. In contrast to linear polarization, the circular polarization allows one to eliminate the Bloch-Siegert shift,[1] increase the effective strength,[2] improve the homogeneity[3] of the field-matter interaction and to address a specific spin state in the case of many-state quantum systems, which is crucial for experiments in the field of quantum information.[4,5]

For small frequencies this task is easily accomplished by two orthogonal coils driven by two radio-frequency (RF) signals phase shifted by 90°, produced by the, so called, quadrature coils. The situation becomes more difficult at microwave (MW) frequencies. In such cases, good results have been obtained by application of two MW strip-line resonators arranged at right angle and fed by appropriately phased MW sources.[4,5,6] Such system works well for a given frequency, however, its high selectivity narrows the tuning range substantially.

In this Letter we present a simple way to produce the circular polarization of the MW (magnetic component) which allows easy addressing of the specific spin state of the investigated sample in a wide range of MW frequencies and magnetic fields.

The designed microstrip circuit can be used in many experiments with various samples like SiV[7], ruby[8], etc., in a bulk or single-center form.

[a)]Electronic mail: mariusz.mrozek@uj.edu.pl

In this paper we describe its application in experiment with ensemble of negatively charged nitrogen vacancy (NV$^-$) color centers in diamond sample interacting with MWs of frequencies between 2.7 GHz and 3.1 GHz, corresponding to resonances in magnetic fields between zero and 100 Gauss.

NV$^-$ color centers in diamond consist of a substitution nitrogen atom and a nearest neighbor vacancy in the carbon lattice. The ground state (GS) of the NV$^-$ ($3A_2$) is a spin-triplet. In the absence of a magnetic field the GS sublevels $m_s = 0$ and $m_s = \pm 1$ are split by 2.87 GHz (Fig.1a) and its gyromagnetic ratio is 2.8 MHz/G. By application of optical pumping (excitation by green light) GS becomes spin-polarized. This allows manipulation of the populations of the $m_s$ sublevels by application of resonant MW field and the optically detected magnetic resonance (ODMR) measurement (Fig. 1). The measurement is based on recording the red fluorescence of the NV$^-$ sample as a function of the MW frequency and can be performed in a range of temperatures, magnetic fields, etc. both with ensembles and single NVs.[9-19] Our optical setup uses confocal imaging (spatial resolution about 10μm) and is described in Ref. 20.

In most of experiments with NV$^-$ diamonds the MW field is linearly polarized. Consequently, in the ODMR experiments with NVs in low magnetic field the resonant linearly polarized MW addresses both transitions (Fig.1a). As shown below, the designed circuit allows experiments where only one of these transitions is preferentially selected in a controlled way.

The applied system for microwave circular polarization consisted of the following elements. The signal was generated by a microwave signal generator (SRS SG386) that had the maximum output power of +10 dBm. It was connected to a 90 degree phase shifter, suitable for use at a few GHz.[21]

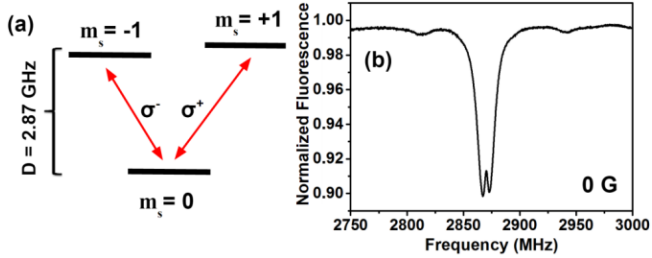

FIG. 1. Level diagram of the ground state of an NV⁻ center (a) and a typical ODMR spectrum (normalized fluorescence vs. the MW frequency) in a zero external magnetic field recorded with a linear polarization of MW (b) (the resonance line is split by crystal strain, the small side resonances come from $^{13}C$, in the center there is a strain-induced dip of NV⁻).

In this device, the input microwave signal is split into two signals shifted by 0 and 90 degrees. To compensate for the phase shift resulting from unequal cable lengths, we used another phase shifter (ARRA 4482D) that allowed us to adjust smoothly the phase $\Delta\varphi$ in one of the channels (Fig. 2).

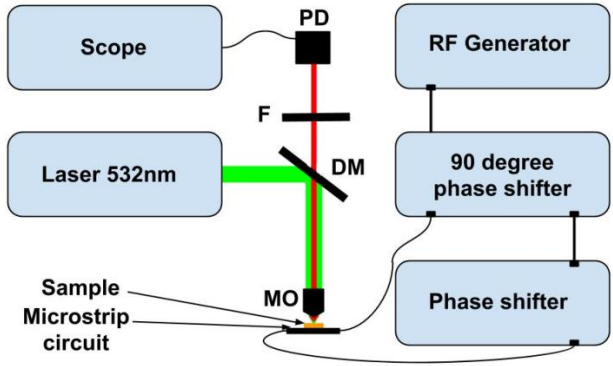

FIG. 2. Experimental setup: PD – photodetector, F – filter, DM- dichroic mirror, MO – microscopic objective.

The two signals were sent to the microwave circuit (Figure 3) with all its outputs terminated in 50 Ω. The circuit enables generation of a strong local MW magnetic field for driving the $m_s = 0 \rightarrow m_s = \pm 1$ transitions at 2.87 GHz. It consists of microstrip lines with a specially designed shape and was fabricated on an insulating substrate (FR4). The bottom layer is covered by the ground plane. The power delivered to the microstrip circuit was between 0 and -2dBm and was symmetrically distributed between the lines.

In the central part of the PCB, the narrow parts of the microstrip lines (2.4 mm long, 0.2 mm wide) are parallel and separated by $d = 1$ mm. When fed by properly impedance-matched MW, they produce strong $H_x$ and $H_z$ components of the magnetic field.

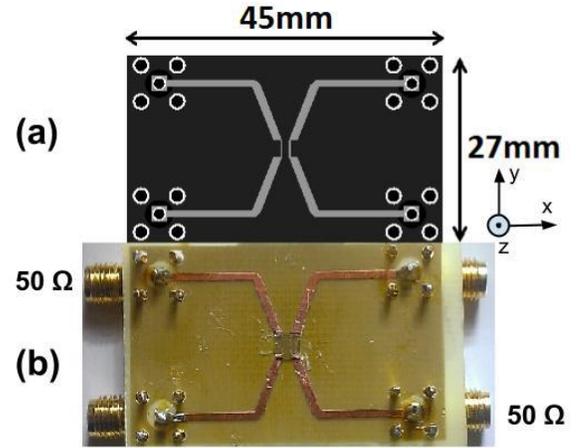

FIG. 3. (a) Layout of the described microwave circuit used for creation of circularly polarized MW. (b) Photograph of the microstrip circuit with the sample placed in the central part. The reverse (bottom) layer is grounded.

Symmetrically between the lines, at the height of $d/2$ above (or below) the plane L1-L1' and L2-L2', $H_x$ and $H_z$ combine to the resulting MW field which becomes fully circular for $\Delta\varphi = \pm 90°$ (Fig. 4).

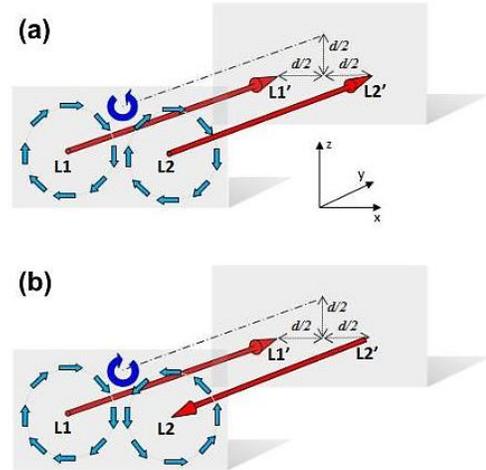

FIG. 4. Geometrical conditions for creation of a circularly polarized MW field with the left-handed (a) and right-handed (b) circular polarization. The parallel lines L1-L1' and L2-L2', separated by $d$ conduct MW currents (red arrows) oscillating with controlled phases ($\Delta\varphi = \pm 90°$) and create circular magnetic MW fields (light blue arrows) in a given (shaded) plane.

Figure 5 presents the simulation of the magnetic field distribution for various relative phase shifts $\Delta\varphi$ between currents in both lines at 0.5 mm above the upper part of the PCB. The sample is placed on the top of the narrow part of the microstrip lines, in the area where the magnetic field is the strongest.

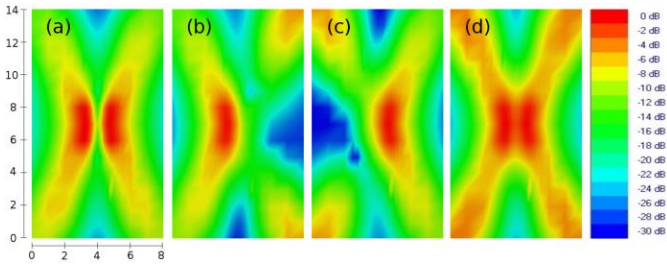

FIG. 5. Simulated magnetic field distribution 0.5 mm above the top layer of the PCB for various values of the relative phase shift between both lines; (a) $\Delta\varphi = 0$ resulting in linear polarization; (b, c) $\Delta\varphi = \pm 90^{o}$ resulting in the left- or right circular polarization. Two time instants are shown corresponding to the maximum strength of the sinusoidal MW signal in the left (b) and right (c) line; (d) time-averaged map of the total intensity for $\Delta\varphi = 90^{o}$. The scale on the axes is in millimeters. The magnetic field (absolute value) is scaled in dB and color-coded (relative to the maximum amplitude (0.8 A/m) for the input power of 0 dBm).

The applied NV$^-$ sample has spatial dimensions of 3x3x0.5mm. The distance $d$ between the microstrip lines must be matched to the distance from the lines to the excitation point. Therefore, the studied point is at $d/2=0.5$mm above the circuit, on the crystal's top surface.

The first measurements were done in a low magnetic field (15 G) for the direction of the magnetic field along the [100] crystal axis [Fig. 6(b)] which results in the simplest, two-component ODMR spectra. By tuning the phase shifter at one of the inputs, we were able to obtain any polarization state of the net MW field, in particular the right-handed ($\sigma$+), left-handed ($\sigma$-) circular polarization or the linear one.

Figure 6(a, b, c) shows the ODMR spectra recorded at magnetic fields of 0, 15 G, and 95 G for three different MW polarizations. The two resonances corresponding to the $m_s = \pm 1$ transitions split in the increasing magnetic field. The created polarization state is frequency- and magnetic-field-dependent, hence the phase shifter needs to be adjusted to maintain a given polarization state when changing either of these two parameters. The observed contrast ratio between the two dips produced by the left and right circular polarizations of the microwaves reflects the purity of the created spin polarization, $p=(a+ − a–)/(a+ + a–)$, with a± being the amplitude of the ODMR component associated with the ±1 transition, respectively.

We have experimentally verified that different distances $d$ between the lines (0.5 and 1.5 mm) resulted in a significantly lower polarization purity than in the optimal case of $d = 1$ mm, which proves the importance of the proper matching of the line and sample dimensions (Fig.4). As discussed in Ref. 4, the maximum available value of $p$ is 93%. At low field, we measured $p$ about 85%, i.e. even higher than in the case of narrowband orthogonal resonators.[6] With increasing intensity of the magnetic field $p$ decreases [Fig. 6(d)], yet even for magnetic fields of about 100 G, it remains greater than 67%. The observed drop of $p$ with magnetic field is caused by the magnetic mixing between other differently oriented NV$^-$ components, discussed in Ref. 22, rather than to any performance reduction of our broadband microstrip line.

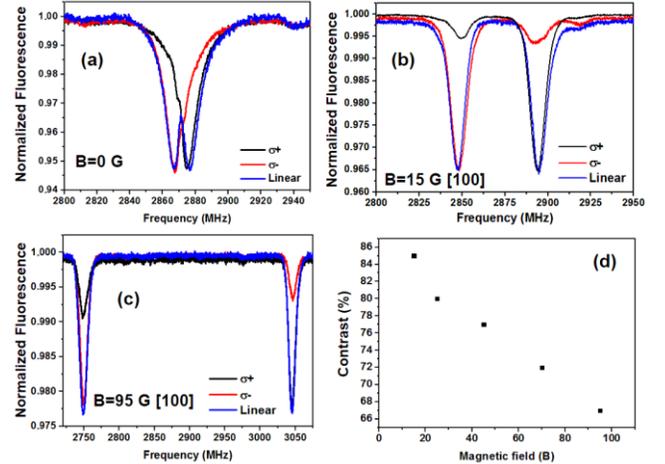

FIG. 6. ODMR spectra at different magnetic field intensities: (a) 0 G; (b) 15 G in the [100] direction; (c) 95 G in the [100] direction, linear polarization signal (blue) was shifted down for clarity.(d) Polarization purity between the two ODMR peaks (addressing the $m_s = \pm 1$ states) versus magnetic field.

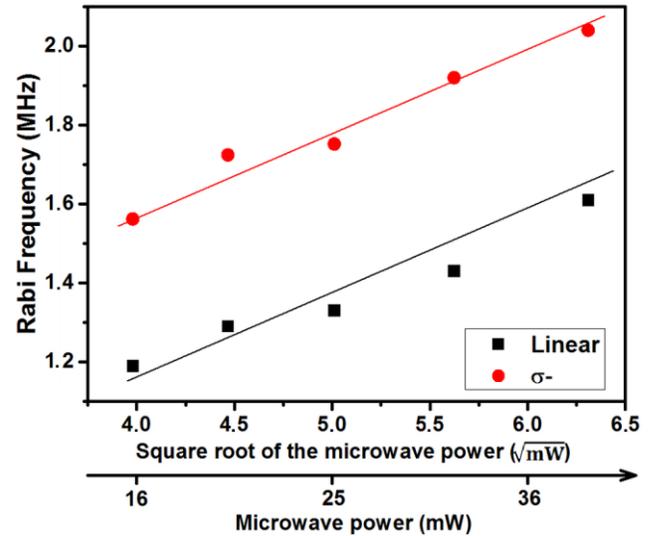

FIG. 7. Rabi frequency vs microwave power for different polarizations. Magnetic field was 15 G in the [100] direction.

One basic method for diagnostics of the spin state of NV$^-$ samples is by recording Rabi oscillations when strongly driving the transition between spin states of interest. Figure 7 depicts how the frequency of the Rabi oscillations (see Ref. 21) for details) measured with our circuit depend on the MW power for different microwave polarizations. The measurement was made at the field of 15 G directed along the [100] axis. As can be seen in Fig. 7, observation of the Rabi oscillations at a given frequency requires twice as high MW power for the linear polarization than for the circular one.

We have presented a simple design of a MW circuit for creation of arbitrary polarized MW magnetic field in a wide frequency range around 3 GHz with well controlled polarization, arbitrarily adjusted between fully linear and circular. We applied it to studies of an ensemble of NV$^-$ color centers in a bulk diamond crystal. By using spin polarization selection rules[5] we were able to address the ODMR resonances in the NV$^-$ sample and drive Rabi oscillations with a controlled MW polarization in the magnetic fields up to 100 G. We have shown that with the circular polarization the NV$^-$ resonances can be studied with the power two times weaker than in the case of linear polarization. This factor of two agrees exactly with the theoretical expectation for pure polarization. Besides the described NV$^-$ measurement, the circuit can be used in a variety of magnetic resonance experiments.

This work was supported by the 7150/E-338/M/2014 (MNSW) and 2012/07/B/ST2/00251 (NCN) grants.